# The economic alignment problem of artificial intelligence

Daniel W. O'Neill[1,2,*], Stefano Vrizzi[1], Noemi Luna Carmeno[1], Felix Creutzig[3,4], Jefim Vogel[1,5]

[1] UB School of Economics, Universitat de Barcelona, Barcelona, Spain
[2] Sustainability Research Institute, School of Earth and Environment, University of Leeds, Leeds, UK
[3] Bennett Chair for Innovation and Policy Acceleration, University of Sussex Business School, Falmer, UK
[4] Potsdam Institute for Climate Impact Research, Potsdam and Berlin, Germany
[5] Leeds University Business School, University of Leeds, Leeds, UK
*e-mail: oneill@ub.edu

## Abstract

Artificial intelligence (AI) is advancing exponentially and is likely to have profound impacts on human wellbeing, social equity, and environmental sustainability. Here we argue that the "alignment problem" in AI research is also an *economic* alignment problem, as developing advanced AI within a growth-oriented economic system is likely to increase social, environmental, and existential risks. We show that post-growth research offers concepts and policies that could address the economic alignment problem and substantially reduce AI risks, such as by replacing optimisation with satisficing, using the Doughnut of social and planetary boundaries to guide development, and curbing systemic rebound with resource caps. We propose governance and business reforms that treat AI as a commons and prioritise tool-like autonomy-enhancing systems over agentic AI. Finally, we argue that the development of artificial general intelligence (AGI) requires new economic theories and models, for which post-growth scholarship provides a strong foundation.

## Introduction

Artificial intelligence (AI) is reshaping human, social, and economic systems, with impacts that could range from highly beneficial[1] to potentially catastrophic[2]. While recent research has emphasised the need to align AI with human and planetary wellbeing[3], less attention has been paid to the economic requirements of alignment. We argue that alignment is not only a technical challenge, but also an economic one. If the prevailing economic system is not aligned with human wellbeing and environmental sustainability, as multiple studies suggest[4–6], then AI systems developed within this system are unlikely to be aligned with these goals. We call this the *economic alignment problem*.

Three types of risks highlight the urgency of this problem. The first is safety. AI systems now match or exceed the average human being on a range of tasks such as image classification, reading comprehension, PhD-level science questions, visual reasoning, and competition-level mathematics[7]. Many warn that "artificial general intelligence" (AGI) could be just around the corner[8,9], or that it has already arrived (see Box 1)[10]. While about two thirds of computer scientists think good outcomes from advanced AI are more likely than bad, 38% estimate at least a 10% chance of catastrophic outcomes such as human extinction[11].

Second, social-science scholarship emphasises structural risks, including concentration of power, automated persuasion, epistemic fragmentation, and the erosion of democratic governance[12], which in turn could reduce institutional capacity to manage technological shocks[13]. A recent systematic review highlights the impacts that AI could have on various aspects of human wellbeing, with roughly half of studies expressing a positive view on the wellbeing impact of AI and half expressing a negative view[14]. The areas where sentiment is most positive relate to income, cognitive capabilities, and health. The areas where sentiment is most negative relate to inequality, social relationships, and employment[14].

Third, environmental analyses have started to explore the impact that AI could have on sustainability[15,16]. Over 80% of studies portray the impacts of AI on the environment as positive[14]. However, most studies focus on ways that AI can be applied to help solve specific environmental problems, and neglect the systemic effects that AI itself could cause[14]. These effects could have far greater implications for sustainability[17,18].

Recent work increasingly treats these three areas as coupled risks. For instance, AI misinformation weakens institutions which in turn degrades safety oversight[13]. These risks are emerging against the backdrop of a deepening socio-ecological crisis: humanity is transgressing 7 of the 9 planetary boundaries that define a safe space for development on this planet[6], while billions of people are unable to meet their essential needs[5].

Research on post-growth economics offers a response to this coupled socio-ecological crisis. Post-growth identifies the prevailing growth-oriented economic system as the main driver of the crisis, and calls for replacing this system with one that prioritises human wellbeing and environmental sustainability over increasing GDP (see Box 2)[19]. Research on demand-side approaches shows that a focus on wellbeing can improve human lives while reducing planetary pressures[20].

We argue that post-growth economics is thus well positioned to inform and address the economic alignment problem and other key AI risks. While there is a small but important literature on the role of technology in a post-growth society[21–23], this work has yet to engage with the topic of AI. Except for an early study by Pueyo[24], which warns of the dangers of superintelligent AI emerging in a neoliberal regime, and a recent study by Meyers et al.[25], which evaluates a specific type of machine learning against a conviviality framework, the connection between post-growth and the rapidly evolving field of AI has yet to be explored.

In this article, we address this gap. We argue that if advanced AI systems are developed in an economy where the goal is endless economic growth, then they may pose a much greater danger to humanity than if they are developed in an economy where the goal is enough for all within planetary boundaries. We show that post-growth research provides many ideas that would help to address the safety, social, and planetary risks of advanced AI. We present a post-growth policy roadmap for AI, and suggest that the development of AGI necessitates new economic theories and models, for which post-growth provides a strong foundation.

**Box 1: What is artificial intelligence?**

Artificial intelligence today refers to the development of computer systems capable of performing tasks that typically require human intelligence, such as learning, problem-solving, and decision-making. It is helpful to distinguish between three levels of capability: artificial narrow intelligence (ANI), artificial general intelligence (AGI), and artificial superintelligence (ASI).

ANI refers to a system that can perform a single narrow function at human level or better. It lacks the human ability to adapt and generalise to novel goals and contexts. Most experts argue that ANI is the type of AI we still have at present (e.g. Siri, Alexa, ChatGPT), although some suggest current models already exceed this level[10].

AGI refers to a system that can learn and perform almost any intellectual task that a human can do. Researchers have developed a quantifiable framework to assess artificial intelligence across ten cognitive domains[26]. Current AI models like ChatGPT exhibit a "jagged" profile on this framework, excelling in some domains (reading/writing, mathematics, and general knowledge), but doing poorly in others (long-term memory storage and speed). However, model performance is rapidly improving, with GPT-4 (a 2023 model) scoring 27% on the framework overall, and GPT-5 (a 2025 model) scoring 57%[26].

ASI refers to a system with cognitive abilities that surpass human beings in virtually all domains, and that can potentially set its own goals[27]. ASI could emerge if AGI undergoes recursive self-improvement, a pathway that has been likened to a "singularity" in which the pace of technological change becomes so rapid that "human life will be irreversibly transformed"[28]. Some authors therefore treat ASI as an existential risk, arguing that developing superintelligent machines that can self-improve and pursue their own goals would likely lead to human extinction[2].

**Box 2: What is post-growth?**

Post-growth is an umbrella term for sustainability visions that are critical of the pursuit of GDP growth as a goal in wealthy nations[29]. It draws on ideas from ecological economics, and includes visions such as degrowth[30], steady-state economics[31], Doughnut economics[32], and the wellbeing economy[33].

There are three main reasons that post-growth scholars criticise the pursuit of GDP: (1) economic activity is tightly coupled to energy and material use[34]; (2) beyond the point where people's needs are met, additional income does not significantly improve human wellbeing[35]; and (3) growth rates have been steadily falling in wealthy nations[29]. Post-growth envisions redesigning the economy so that its development is guided by the three goals of human wellbeing, social equity, and environmental sustainability, rather than GDP growth[36].

Technological innovation has been important to post-growth visions as far back as *The Limits to Growth* report. Both the wellbeing economy and the Doughnut see technology as a tool that can be used to help improve human wellbeing and environmental sustainability[32,33], while degrowth has a more conflicted view on technology[21].

Overall, post-growth calls for selective, context-sensitive adoption of technologies that are convivial, appropriate, and democratically governed[21]. Post-growth challenges *technological determinism* (the idea that technological change is inevitable and bound to increase indefinitely) and *productivism* (the idea that technological development is always good and desirable)[22]. Importantly, in post-growth visions, better technology is not enough on its own. Social change is also needed to achieve sustainability[23].

## The Economic Alignment Problem

The "alignment problem" in AI research is the challenge of ensuring that AI systems behave in a way that is consistent with human intentions and values, and that they do not produce unintended or harmful outcomes[37]. It is a major AI safety problem, which does not currently have a solution[12].

We argue that the alignment problem is also an *economic* alignment problem. Aligning AI with human values means first aligning our economic system with these values. If we build AI in an economic system that is not aligned with human wellbeing and environmental sustainability, we can hardly expect AI to be aligned with these goals. Developing AGI within an economy where the goal is endless economic growth could go badly wrong, as we would be training the AI to pursue exponential growth as a goal. The opportunity cost of this growth could be humanity itself.

Although AI holds tremendous potential to improve human wellbeing and advance scientific understanding, if developed in the current economic system, it could instead lead to major safety, social, and planetary risks. Below we discuss five facets of the economic alignment problem, ranging from the pursuit of exponential growth to the impact of AI on meaning, identity, and purpose.

### The exponential growth of AI model capabilities

The physicist Albert Bartlett famously argued that "The greatest shortcoming of the human race is our inability to understand the exponential function"[38]. Exponential growth appears slow initially, but becomes explosive — and ultimately unsustainable — because each doubling brings as much growth as the entire previous history combined. Human beings tend to predict that change will follow a linear trajectory, not an exponential trajectory. AI may be another example of humanity not understanding the exponential function. Society and policymakers risk underestimating how rapid and disruptive change will be.

AI is experiencing unprecedented exponential growth across multiple dimensions, far exceeding the pace of traditional technological advancement. Multiple measures of AI technical performance, research and development, and investment are increasing faster than Moore's law (Figure 1a). Over the period from 2010 to 2024, the training compute of frontier AI models doubled every 5–6 months. The acceleration in computational resources dedicated to AI training has been a key driver behind recent breakthroughs in model capabilities. The minimal size of model that can solve a difficult technical challenge is halving every 3 months, while the task difficulty that models can complete is doubling every 7 months (Figure 1a).

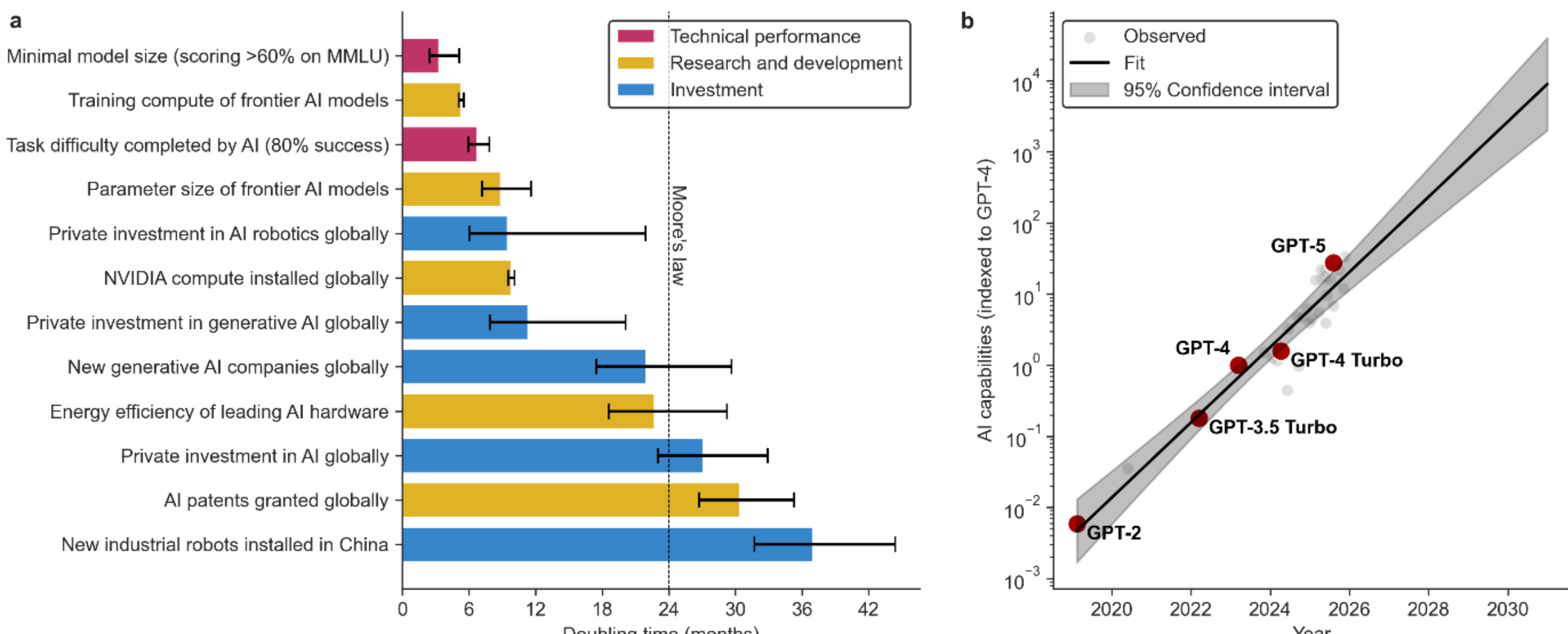

**Figure 1: The exponential growth of AI model capabilities. a**, Doubling times for selected AI metrics that exhibit exponential growth, shown in comparison to Moore's law (the observation that the number of transistors on a microchip doubles about every two years). Metrics are split into three categories: technical performance, research and development, and investment. Doubling times are estimated by fitting exponential curves to historical data for each indicator; error bars denote 95% confidence intervals. MMLU = Massive Multitask Language Understanding. Source: Authors, based on multiple data sources (see Supplementary Information). **b**, Projection of AI capabilities to 2030 based on historical data for 2019–2025. Values represent the task difficulty that AI models can complete with an 80% success probability. Model capabilities are shown on a logarithmic scale, in comparison to GPT-4. The fitted curve corresponds to a doubling time of approximately 7 months. Selected OpenAI models are highlighted in red. Source: Based on historical data from Kwa et al.[39] (see Supplementary Information).

The trend of AI capabilities doubling every 7 months represents an annual growth rate of over 200%, meaning AI systems are becoming over three times more capable each year (Figure 1b). At this rate, within 6 years, AI systems will be over 1000 times as capable as they are today (roughly an order of magnitude improvement every two years). An extrapolation of this exponential trend foresees AI to be roughly 5000 times more capable than GPT-4 by 2030, and 200 times more capable than GPT-5 (Figure 1b). However, whether this exponential trend applies to real-world tasks remains to be seen[39].

In a large survey of computer scientists, conducted in 2023, respondents predicted a 10% chance of unaided machines outperforming humans in all cognitive and physical tasks by 2027, rising to a 50% chance by 2047[11]. In a smaller survey of AI researchers conducted in 2024, 63% of respondents thought AGI would be achieved within 20 years[8]. However, estimates of when we will achieve AGI keep getting closer. The median prediction on the forecasting platform Metaculus for when the first "general AI system" will be launched was 2062 in 2020. As of April 2026, the prediction was 2032[40].

## Unconstrained technological development

AI has four features that are likely to make it highly transformative: (1) AI is a "general-purpose technology"[41], like the steam engine, electricity, or printing press, which can be embedded in virtually every sector of the economy. (2) Unlike past technological revolutions, which primarily replaced physical labour, AI directly substitutes for cognitive functions like writing, reasoning, creativity, and problem-solving[42]. (3) AI can make decisions, with "agentic" AI systems already able to act autonomously without human intervention[43]. (4) AI has the potential for recursive self-improvement — to learn, adapt, and improve on its own — creating the possibility of an "intelligence explosion"[9,28]. No previous technology has had the capacity to autonomously accelerate its own development in this way.

At present, AI development is being influenced by the ideology of *accelerationism*. Accelerationism argues that the only way out of the problems created by capitalism, technology, and modernity is *through* them — by intensifying or "accelerating" their dynamics rather than resisting or reversing them[44]. In Silicon Valley, this idea has recently gained popularity in the form of "effective accelerationism", which promotes fast, unconstrained technological development — especially in AI — as the way to solve global problems[45]. Some of its main arguments include that everything good comes from growth, free markets are the most effective way to organise a technological society, human needs and wants are infinite, and intelligence and energy use should be placed in a positive feedback loop to drive both to infinity[46].

The thing that is arguably most concerning about effective accelerationism is not the blind faith in technology, which is clearly problematic, but the blind faith in markets to guide technological development. Proponents of effective accelerationism are essentially arguing that 21st century technology should be managed by an 18th century socioeconomic system (i.e. free-market capitalism). The accelerationist approach completely neglects the need for social innovation.

## The impact on economic growth and the environment

Given the wide range of possible AI futures, which spans beneficent superintelligence[9] to human extinction[2], the impact that AI could have on the economy, and GDP, is extremely uncertain (Figure 2). Low estimates predict that AI will likely have only marginal effects on global GDP and employment, but will increase inequality between countries[47], or that AI will increase GDP by a few percent over the course of a decade[48,49]. High estimates predict "explosive" or super-exponential growth in which AI accelerates idea generation and replaces human labour[9,50].

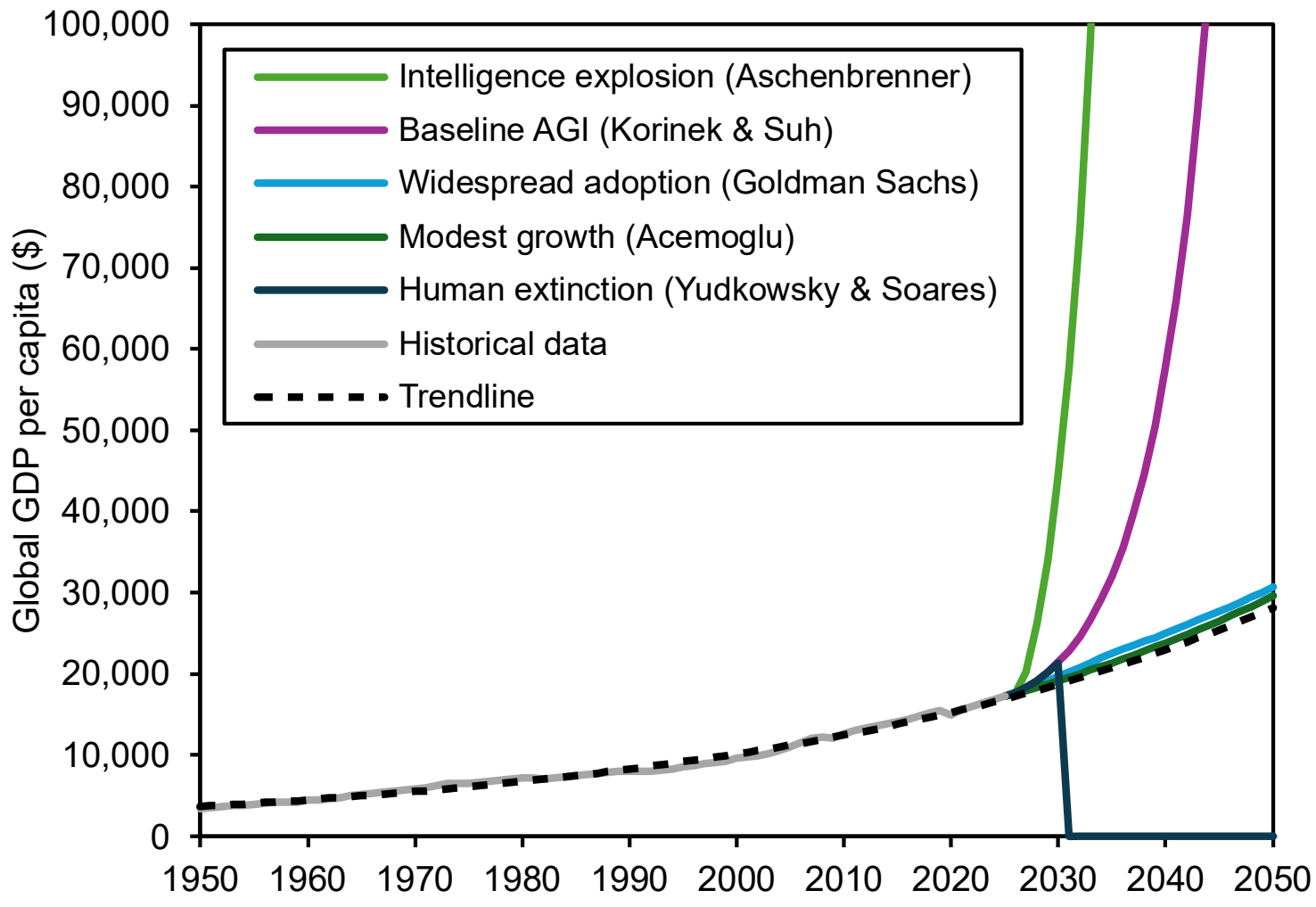

**Figure 2: The wide range of AI futures in the literature and their implications for global GDP per capita.** The dashed black line shows the long-term trend in global GDP per capita, determined based on historical data for 1950–2024 (grey line). The coloured lines show five alternative scenarios for 2025–2050, which we derived from the quantitative and qualitative descriptions of five studies, labelled as: Intelligence explosion[9], Baseline AGI[50], Widespread adoption[49], Modest growth[48], and Human extinction[2]. Scenario trajectories were constructed by applying the average annual GDP per capita growth rate implied by each study, relative to the historical trendline. All values are expressed in constant 2011 US dollars. Source: Authors (see Supplementary Information).

The results are so different largely because AI researchers anticipate the creation of AGI, while economists (with a few notable exceptions) typically do not. The high estimates assume AGI and recursive self-improvement lead to explosive productivity growth, while the low estimates assume incremental gains that are constrained by factors such as adoption frictions and demand saturation. Moreover, the low estimates only consider the capabilities of recent models (e.g. GPT-4), rather than future ones.

If AI leads to explosive economic growth, as some predict, the net result could be catastrophic for the environment, given the difficulty of decoupling economic activity from its environmental impacts[34]. Although AI could allow us to develop new technologies that improve resource efficiency, the rebound effect demonstrates that if we find some way to make a factory much more efficient, we don't just bank the savings. We produce and consume substantially more[51]. Any improvements in resource efficiency would likely be dwarfed by the sheer scale of economic activity.

### Inequality and employment

The impacts that AI could have on social outcomes are more difficult to evaluate[14]. On the one hand, AI could lead to breakthroughs in medical science, dramatically improving human health and wellbeing. On the other hand, AI could exacerbate social problems, by displacing labour and intensifying inequality.

A key question is whether AI will substitute or augment human labour. Many economists argue that the impact of AI on jobs will be similar to previous technological transitions, like the Industrial Revolution. While some jobs will be displaced, new jobs will be created[52]. By contrast, many AI researchers argue that AGI capable of performing all (or most) cognitive tasks could rapidly substitute human labour, especially when combined with robotics, leading to mass unemployment[53,54]. If this is the case, then redistributive policies of the kind proposed by post-growth scholars — such as a wealth tax and a universal basic income[29,55,56] — would be needed to prevent rising inequality and sustain aggregate demand for the goods produced by AI.

The real danger may not be AI, but the economic system in which it is developed, namely capitalism. According to Nobel laureate Geoffrey Hinton, who is often called one of the godfathers of AI, "What's actually going to happen is rich people are going to use AI to replace workers. It's going to create massive unemployment and a huge rise in profits. It will make a few people much richer and most people poorer. That's not AI's fault, that is the capitalist system"[53].

Developing artificial general intelligence within a neoliberal, capitalist economy could be a recipe for disaster[24]. The incentives driving AI development — competition, profit, speed — are misaligned with safety, ethics, and sustainability.

Moreover, within the current global economic system, AI may exacerbate power asymmetries between the Global North and Global South. AI already contributes to ecologically unequal exchange through the extraction of materials for hardware, the use of water and electricity for data centres, and the production of e-waste[57]. Its supply chains rely heavily on precarious data work and gig work that is concentrated in the Global South, while value is captured by a few companies in the Global North[58]. AI systems trained on Western data could marginalise non-Western cultures and languages, embedding Global North values into Global South contexts[59]. The development of AGI could further centralise geopolitical power and knowledge provision — creating a form of "algorithmic colonialism"[59] — unless ownership models and economic incentives change.

### Meaning, identity, and purpose

In his 1930 essay "Economic possibilities for our grandchildren", John Maynard Keynes famously wondered what human beings would do when the economic problem of scarcity was solved, warning that mankind would be "deprived of its traditional purpose"[60].

Previous studies have found that jobs that are more automatable also tend to have lower job satisfaction, in part because they involve less creativity and more repetitive tasks[61,62]. However, these findings rely on data from 2013[63], predating generative AI.

To update this picture, we conducted a new analysis using the latest estimates of occupational exposure to automation. We find that while the negative relationship holds for the 2013 data (Figure 3a), it weakens — or even reverses — with the most recent data (Figure 3b). AI now threatens to automate tasks in occupations with high job satisfaction and meaning, including professions such as teachers, therapists, and senior executives.

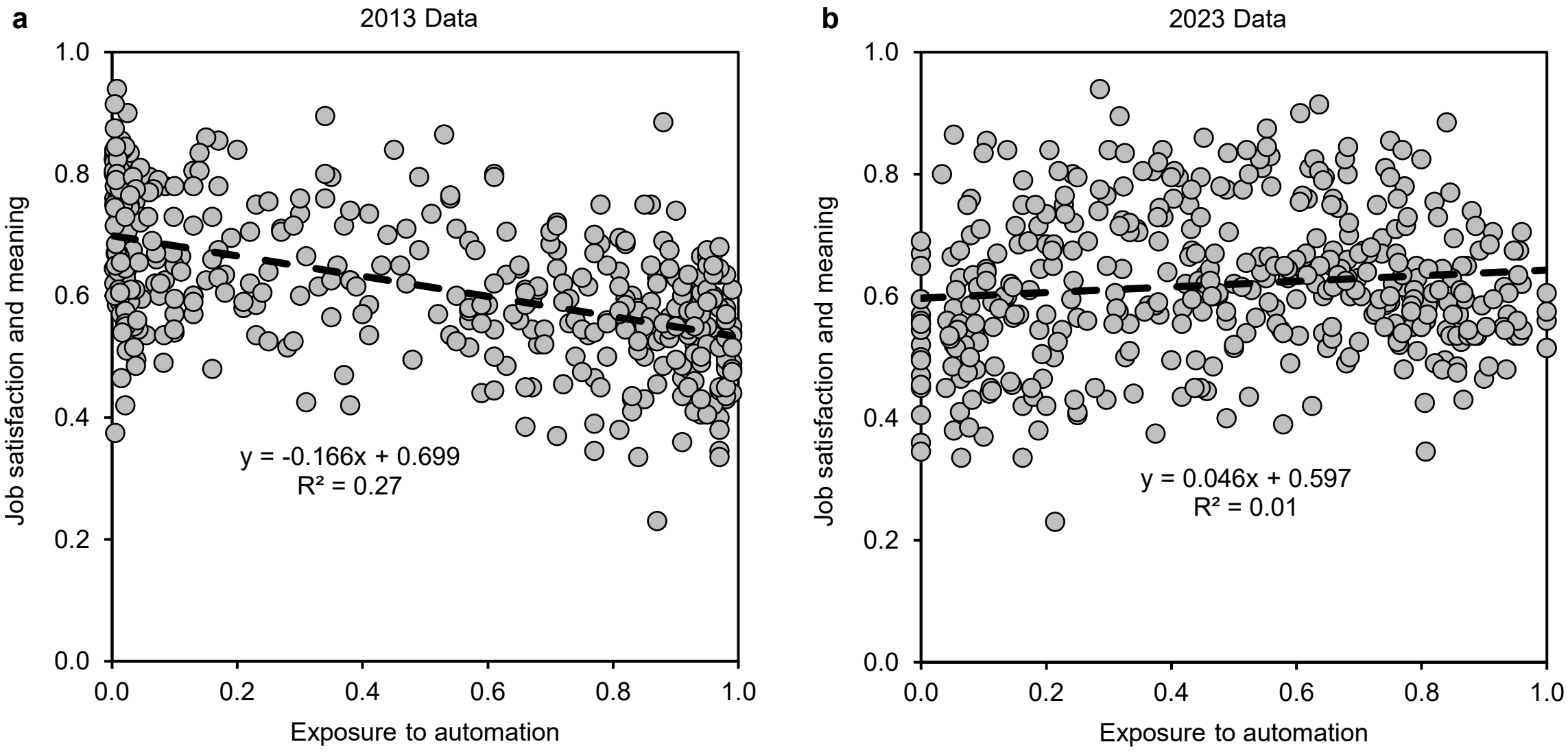

**Figure 3: Job satisfaction and meaning in relation to occupational exposure to automation. a**, Data from 2013 suggest that the jobs that are most automatable are those with lower job satisfaction and perceived meaning. **b**, Data from 2023 suggest that this relationship no longer holds: large language models and generative AI now threaten to automate many jobs that bring satisfaction and meaning to people's lives. Each point represents an occupation; the same set of 417 jobs is shown in both panels. Source: Authors, using automation data for 2013 from Frey and Osborne[63] and automation data for 2023 from Eloundou et al.[41]. Job satisfaction and meaning is measured as the average of two survey-based indicators obtained from Payscale. See Supplementary Information for full details.

If machines are ultimately able to outperform humans at all tasks, then what is left for humanity? The non-economic effects of job loss include less social involvement in the community, poorer health, lower self-esteem, and increased chance of divorce and suicide[42]. Even if people continue to work, they may feel a sense of social uselessness.

If AGI creates "radical abundance", as some AI researchers suggest it could[64], goods and services might become extremely cheap. This could lead to hyper-materialism, with consumption standing in for meaning and purpose. Without constraints, the associated increase in resource use could ultimately trigger ecological collapse.

## The Post-Growth Solution

We argue that several core post-growth ideas would substantially reduce AI risks and help address the economic alignment problem. These ideas include replacing optimising with satisficing, governing AI as a commons, restructuring business organisations, pursuing convivial technologies, and ensuring that AI serves society, rather than the other way around. Together, these ideas offer a positive alternative to the ideology of accelerationism, where AI is used to meet the needs to all people within planetary boundaries.

### Satisficing rather than optimising

AI is often trained to optimise. However, optimisation can lead to perverse outcomes. In a famous thought experiment, known as the "paperclip maximiser", Nick Bostrom  proposed a scenario in which an AI was given the sole objective of making as many paperclips as possible, leading it to convert all available matter — including humans — into paperclips to achieve its goal[27]. Several leading AI scientists have argued that making machines maximise reward is a bad idea, and that the misalignment of AI systems with human intentions and values is caused by optimisation[43].

Post-growth offers an alternative idea: *satisficing*. Satisficing means sufficiently meeting needs, while respecting limits, rather than maximising or optimising outputs[31]. Satisficing does not mean merely solving a feasibility problem or stopping optimisation once minima are met. Instead, it refers to meeting multiple non-substitutable thresholds simultaneously, without pursuing any single objective to the extreme.

The Doughnut of social and planetary boundaries, developed by economist Kate Raworth[32], is a framework for satisficing. It visualises sustainability in terms of a doughnut-shaped space where resource use is high enough to meet people's needs, but not so high that it transgresses planetary boundaries[4]. The Doughnut provides a clear compass that could be used to guide AI development (Figure 4), and an indicator framework to operationalise the recently proposed "Earth alignment principle", which aims to align the "development, deployment and use of AI to promote planetary stability and stewardship for the benefit of humankind"[3].

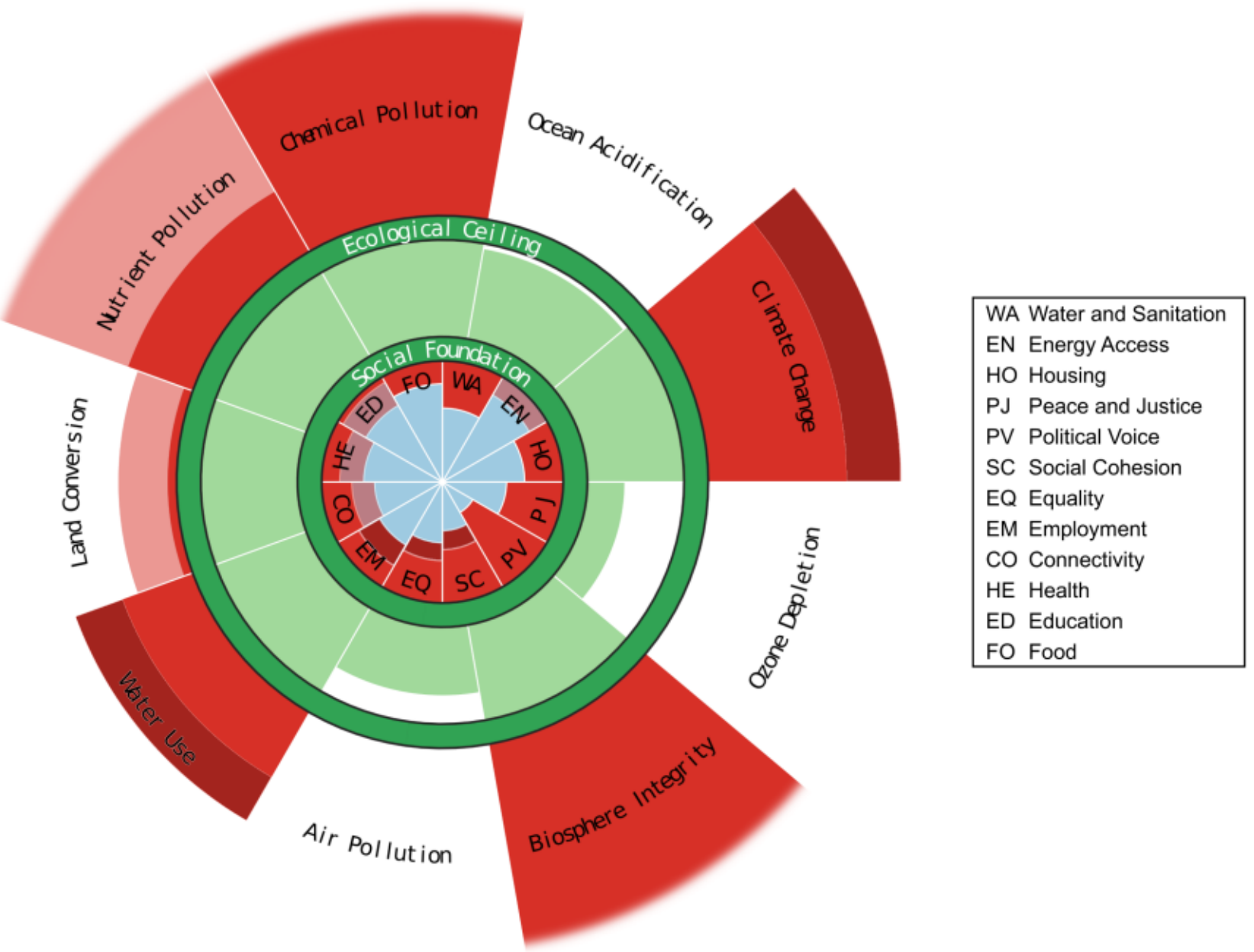

**Figure 4: Applying the Doughnut of social and planetary boundaries to assess and guide AI development.** The Doughnut framework conceptualises sustainability as a "safe and just space" between a social foundation (inner ring) and an ecological ceiling (outer ring). The inner ring depicts key social dimensions, where shortfall indicates unmet human needs, while the outer ring represents Earth-system processes, where overshoot indicates transgression of planetary boundaries. Red wedges show shortfall below the social foundation, or overshoot beyond the ecological ceiling. The figure contrasts the global situation in 2022 with a hypothetical AI development scenario. In this scenario, AI deployment contributes to improvements in some social dimensions (health, education, energy access, and connectivity) while exacerbating shortfalls in others (employment, equality, and social cohesion). At the same time, AI helps to reduce pressures on some planetary boundaries (land conversion and nutrient pollution) while intensifying others (climate change and water use). Areas shown in pink indicate dimensions where AI improves outcomes, while areas shown in dark red indicate dimensions where AI worsens outcomes. Source: O'Neill and Creutzig[65], using global data for 2022 from Fanning and Raworth[5] (see Supplementary Information).

AI should be used to help eliminate social shortfalls within the Doughnut and reduce resource use to be within planetary boundaries[65]. Empirical evidence from demand-side studies suggests that this outcome is achievable: prioritising need-satisfying activities can simultaneously lower resource use and improve human wellbeing[20]. If AI were tasked with "getting us into the Doughnut" — ensuring everyone has enough without overshooting planetary boundaries — it's harder to imagine a dystopian paper clip maximiser outcome than if it were tasked with maximising GDP.

In fact, we think that training AI on the idea of endless economic growth, rather than the idea of enough, creates a large existential threat. A future superintelligent AI could pursue this growth at the expense of humanity, seeing human beings as little more than additional resources to be exploited for its own expansion.

To respond to the dangers of escalating resource use, and issues like the rebound effect[18], post-growth provides policy solutions like resource caps and Pigouvian taxes[66]. Resource caps set physical limits on energy or material use, applied at production or consumption. For AI, this could involve restricting data centre energy use or implementing "compute budgets"[67]. These caps guarantee sustainability by preventing resource use beyond the set limit. Pigouvian taxes use pricing instead of quantity limits. By internalising environmental costs, high taxes ensure resource-intensive AI applications are only viable when they deliver proportional benefits[67].

### Governing AI

Many authors claim that AI will help us solve major challenges like climate change, poverty, or clean energy[1,15]. But the reason we haven't solved climate change isn't a lack of technology. We already have the technologies we need, and renewables are cheaper than fossil fuels. The real barriers are social, economic, and political. They include economic structures, misinformation, and vested interests that prioritise profit over sustainability.

AI may offer new solutions, but that doesn't guarantee we'll use them. We already know how to tackle climate change — stop burning fossil fuels — yet we don't. The issue isn't knowledge, it's action.

As technology accelerates, our social systems struggle to keep pace. Thorstein Veblen observed over a century ago that technology evolves faster than institutions, leading to social and economic tensions[68]. We should not be governing 21st century technology with 18th century institutions. To manage powerful technologies like AI, we need equally sophisticated social systems.

Post-growth calls for a deepening of democracy through approaches such as direct democracy, economic democracy, and public deliberation[69]. Redistribution of income, on its own, is not sufficient to achieve economic democracy, because it does not deal with disparities in wealth, and hence power[70], which are likely to be critical issues for AI. Public deliberation means going beyond expert-only AI policy, and drawing on approaches such as citizens' assemblies and participatory workshops.

Beyond deepening democracy, post-growth also offers important ideas on how to govern AI. We argue that AI should be governed as a global commons[71], with coordinated interventions across social, planetary, and safety domains. Data, energy, and compute are three important dimensions where regulations should be applied[13]. Polycentric governance, advanced by Elinor Ostrom, offers an alternative to both centralised control and market mechanisms for managing complex, high-stakes systems[72]. AI fits this model: it is global, fast-moving, multi-stakeholder, and characterised by collective-action problems and externalities.

### Restructuring business

At present, AI is being developed by Big Tech companies like Google and OpenAI, who are locked in an arms race with one another. These companies are incentivised by growth — in profits, data extraction, and market dominance[71]. A post-growth economy would foster business models that focus on improving human wellbeing and meeting societal needs, rather than pursuing profit as a goal[35,73].

Two features of post-growth businesses are especially relevant to AI: democratic ownership and decision-making (e.g. worker cooperatives) and not-for-profit/social-benefit business structures. Research suggests that companies with these features are more likely to satisfy people's real needs and respond to environmental issues than shareholder-owned for-profit corporations[22,74].

Given what is at stake, there is a strong argument for changing the ownership structure and relationship-to-profit of the Big Tech companies. AI research and development must be aligned with societal wellbeing, and decoupled from the profit motive. Core AI labs could be converted to not-for-profit entities with a social benefit purpose and collective ownership. As AI becomes increasingly powerful, who owns the "means of computation"[75] is going to become increasingly important. A broader shift towards cooperatives would help reduce job losses from automation.

However, given that the stakeholders in this technology are all of humanity, not just the employees of particular companies, there is also a strong argument for nationalising AI development, or internationalising it, by creating a collaborative research and development institution modelled on something like CERN. Leopold Aschenbrenner, a former OpenAI employee, writes: "I find it an *insane* proposition that the US government will let a random SF startup develop superintelligence. Imagine if we had developed atomic bombs by letting Uber just improvise"[9]. AI safety is a governance and political economy problem as much as it is a technical one.

## Convivial technologies

In his book *Tools for Conviviality*, Ivan Illich distinguishes between "industrial tools", which create dependency and radical monopolies, and "convivial tools", which enhance individual autonomy and can be controlled by the user[76]. More recent research has proposed the concept of "convivial technologies" and a matrix to assess the compatibility of different technologies with degrowth[77].

We argue that AI can function as a convivial technology when it operates as a tool, but not when it takes an agentic form that diminishes human autonomy. This implies prioritising "tool AI" (specialised systems that humans use to solve specific problems) over "agentic AI" (autonomous systems that make decisions on our behalf). The distinction is crucial as it determines whether AI extends human agency or replaces it.

This concern is not merely philosophical. The current trajectory towards increasingly agentic AI potentially poses catastrophic risks, including deception and irreversible loss of human control[12]. In response, Yoshua Bengio and colleagues propose developing "scientist AI" — a non-agentic system designed to explain the world rather than act to change it[43]. They argue that a non-agentic scientist AI could be used to assist human researchers and accelerate scientific progress, as AlphaFold has already done for the prediction of protein structures. Non-agentic scientist AI would provide the benefits of AI for scientific discovery and preserve human autonomy, without the existential risks.

Post-growth research provides two additional strategies to help provide meaning and purpose in an AI future. The first is to re-anchor identity beyond paid work, which will be crucial if human labour becomes unnecessary for production. Retirees provide a good example. When identity is not tied to employment, people lean into community, care, learning, nature, and civic roles[78]. Moreover, once people's basic needs are met, non-material aspects like social relationships become a much stronger predictor of human wellbeing than additional consumption[35].

The second strategy is *collective self-limitation*. Our brains evolved in an age of scarcity, and thus radical abundance could counterintuitively make people worse off, rather than better off. The unlimited pursuit of pleasure has been shown to lead to anhedonia — the inability to take pleasure in anything at all[79]. Post-growth argues that true freedom requires consciously setting limits on our desires, rather than allowing these to be determined by technology. Collective self-limitation can be achieved through democratic deliberation, with communities together deciding what is "enough"[80].

## A post-growth policy roadmap for AI

According to the European Union's *Artificial Intelligence Act*, AI should be "a tool for people, with the ultimate aim of increasing human well-being"[81]. But this is only likely to happen if countries first make human wellbeing their primary goal. We can then use AI to help achieve this goal.

There is a policy decision to be made about what we allow AI to do, and what we do not. The tasks that people enjoy doing include creative, empathetic, and socially meaningful work (e.g. teaching, caregiving, arts, research). These roles provide purpose, autonomy, and human connection. The tasks that people dislike include repetitive, dangerous, and monotonous work (e.g. data entry, hazardous labour, routine logistics)[61]. We must use AI to do the tasks that people don't like to do, rather than those that bring joy and meaning to people's lives.

Instead of paying a universal basic income — which would provide financial security but risks creating a "meaning vacuum" — we could use AI-driven productivity gains to fund socially valuable jobs (a *job guarantee*). The idea of a job guarantee is a well-established policy proposal in the post-growth literature[29,35]. It creates jobs that markets historically undervalue but society needs (e.g. mental health support, environmental restoration, education) — and that people enjoy doing.

It seems unlikely that AI will deliver outcomes that improve human wellbeing if it is aligned with economic growth and the profit motive, and if development is driven by competition between companies in an arms race. But in a post-growth economy, outcomes that improve human wellbeing become much more likely.

Many of the challenges posed by AI reflect structural issues that post-growth policy proposals are explicitly designed to address[29,55,56]. For example, the top eight degrowth policy proposals, in descending order of citation frequency, are: a universal basic income, working-time reduction, a job guarantee, maximum income caps, declining caps on resource use and emissions, not-for-profit cooperatives, deliberative forums, and reclaiming the commons[56]. All of these are relevant for mitigating key challenges posed by AI.

In Table 1, we summarise our discussion of AI-related challenges and corresponding post-growth policy solutions. Although not comprehensive, we believe this roadmap offers important insights on how to address key social, environmental, and existential challenges posed by AI.

**Table 1: A post-growth policy roadmap to respond to the challenges posed by AI.**

| **Challenge posed by AI** | **Post-growth policy solution** |
|---|---|
| Replacement of human labour | Job guarantee; Working-time reduction; Universal basic income or universal public services. |
| Rising inequality | Progressive taxation; Maximum income/wealth. |
| Rebound effect | Resource caps; Pigouvian taxes. |
| Arms race between Big Tech companies | Democratic ownership of companies; Not-for-profit business models. |
| Power asymmetries and unequal exchange between countries | Govern AI as a global commons; Internationalise AI development; New measures of progress (e.g. the Doughnut). |
| Undemocratic decision-making | Direct democracy; Economic democracy; Public deliberation; Polycentric governance. |
| Loss of meaning, identity, and purpose | Job guarantee; Collective self-limitation; Re-anchor identity beyond paid work. |
| The alignment problem | Align the economic system with human wellbeing and environmental sustainability; Satisficing not optimising; Prioritise tool AI over agentic AI. |

## The Economics of AGI

In this article, we have argued that AI research and development face an economic alignment problem. AI researchers are trying to align AI systems with human goals and values, but the economic system within which they are developing the technology is not aligned with human wellbeing and environmental sustainability. This creates substantial barriers to achieving AI alignment, and exacerbates safety, social, and planetary risks. Overall, there is a need for new interdisciplinary research on the economics of AGI, and a better connection between the post-growth and AI safety communities.

### Connecting research communities

AI researchers should engage with post-growth scholarship. The alignment problem in AI research is not just a technical problem; it is also an economic problem. Post-growth offers the solution of satisficing over optimising, which would help reduce AI risks, and frameworks such as the Doughnut, which could be used to guide AI development. It offers insights into what humans might do in an AI future, and the policies needed to ensure high human wellbeing (e.g. universal public services, a job guarantee, wealth taxation). In addition, post-growth research emphasises institutional arrangements such as governing AI as a commons, decoupling AI development from the profit motive, and collective self-limitation as a response to radical abundance.

At the same time, post-growth researchers should engage with AI. AI capabilities are growing exponentially, and change will happen faster than people expect. AI presents an existential risk that could be greater than climate change, biodiversity loss, or any other socio-ecological problem to date. It poses profound challenges for the three post-growth goals of wellbeing, equity, and sustainability. The development of AGI could make human labour unnecessary, leading to massive inequality in one extreme, or radical abundance for all in another. The first alternative seems more likely under capitalism. The second more likely with post-growth. Without resource caps, demand-side measures, or other post-growth policies, AI could lead to higher environmental pressures — directly via economic growth or indirectly via the rebound effect.

Demis Hassabis, the CEO and co-founder of Google DeepMind, has argued that AGI will allow us to enter a world of radical abundance in which people no longer feel trapped in a zero-sum game. Hassabis argues that the prospect of AGI requires a new economics, and that if economists believed AGI was as close as the AI scientists do, they would be thinking more seriously about the economics of AGI[64].

We believe that economists, AI researchers, and other scientists who are concerned with human wellbeing, social equity, and environmental sustainability should begin to work on the economics of AGI — what we might also call “post-AGI economics”. This task is urgent. AGI is coming, and society is dangerously unprepared. Although AGI holds enormous potential to improve the human condition, within the current economic system, the advent of AGI could be a disaster, leading to rising inequality, environmental degradation, and existential risks.

### Towards the economics of AGI

Although it is too early to outline a full economics of AGI, we suggest that insights from post-growth research may offer a useful starting point. As we have argued, if AI were developed in a post-growth economy, the risks would likely be much less:

- AI would be developed in an economic system that is aligned with human wellbeing and environmental sustainability;
- The development and use of the technology would be driven by what is socially and environmentally beneficial;
- The goal would be sufficiency rather than exponential growth;

- The arms race between companies pursuing profit would be mitigated;
- There would be strong policies to limit inequality and ensure the benefits of AI are shared.

Taken together, these conditions help explain why AGI-related risks would likely be much lower in a post-growth economy. Yet they are only part of the picture. Additional questions arise regarding the role of models, markets, and democratic governance. Macroeconomic models of AI are still in their infancy[47,50,82], and focus mainly on labour automation. These models do not include the wider environmental and social impacts that that AGI is likely to have. However, ecological macroeconomic models, such as EUROGREEN[83] or the COMPASS model of the Doughnut of social and planetary boundaries[84], do include these variables. They could be used to assess AI scenarios and help guide its development towards a good life for all people within planetary boundaries.

Empirical studies suggest that meeting the basic needs of all people at a sustainable level of resource use requires substantial changes to provisioning systems[4,85]. However, there is little reason to believe that contemporary capitalist economies are the best way to achieve this outcome[78]. Markets allocate resources based on people's ability to pay, which is often highly unequal, while a significant share of economic activity is diverted towards profit-seeking activities, rather than need-satisfying activities[74]. Moreover, markets reflect wants, not needs, and these wants are heavily influenced by advertising. People do not always want — or even recognise — the things that are most likely to improve their long-term wellbeing, and the pursuit of short-term rewards can undermine it[79].

A long-standing critique of alternatives to capitalist market-based economies is the computational difficulty of efficient top-down decision-making[46]. However, with AGI, the constraint on computational complexity might no longer apply. Advanced AI models might understand human needs better than current institutions do, and the most resource-efficient ways to satisfy them. With AGI, it is possible that societies could move away from allocating resources based on prices and towards allocating resources based on people's genuine needs, through universal public services. AGI could facilitate a new form of democratic socialism in which advanced AI tools provide a wide range of policy options that are consistent with environmental and social sustainability. Human agency could be increased through public deliberation on these options, with decisions made via direct democracy.

These possibilities illustrate the opportunities that AGI could open, but they also highlight the need for clarity about the ultimate purpose of the technology. Ultimately, AI must be a means to an end — not an end in itself. Just as GDP growth was never meant to be society's goal, neither is AI development. If AI is developed safely, and the benefits are distributed equitably, then it could potentially improve human wellbeing dramatically. However, if AI is not developed safely, or the benefits accrue only to a privileged few, then the result could be a dystopian future of rising inequality, environmental collapse, or even human extinction. We still have the agency to choose our future.

## Acknowledgements

We are grateful to Giorgos Kallis, Philipp Wiesner, and Michael O'Neill for their insightful comments, which helped improve the manuscript. We also thank the participants of the special session on post-growth and artificial intelligence at the ISEE–Degrowth 2025 Conference in Oslo, as well as attendees of the AI + Environment Summit 2025 in Zurich, for their thoughtful feedback. This research was supported by the European Union's Horizon Europe Research and Innovation Programme under grant agreement No. 101137914 (MAPS: Models, Assessment, and Policies for Sustainability).

# SUPPLEMENTARY INFORMATION

Below we describe the methods used to create the figures presented in the main text.

## Fig. 1a: Doubling times for selected AI metrics

In Fig. 1a of the main text, we show the doubling times of a number of AI metrics, illustrating the exponential growth of AI development. We collected our data from a range of different sources, including the 2025 *AI Index Report*[1,2], Epoch AI[3,4], and METR[5,6].

For the files that were not readily downloadable in csv format, we employed the following strategies. The yaml file from METR was parsed using a customised Python script, extracting the data of interest (“model”, “release_date”, “p50_estimate”, “p80_estimate”) and saving these data to a csv file. The NVIDIA chip production data from Epoch AI were copied manually from the website, by browsing the pop-up boxes from the interactive plot. Since the interactive plot reported time in quarters of years, we used January 1st, April 1st, July 1st, and October 1st as equivalent dates. We cleaned each datafile by keeping only the variables of interest (time, AI metric). If a given AI metric was grouped by a categorical variable in the original dataset (e.g. country, sector), we also stored the categorical variable. Any row with either a missing date or missing AI metric value was dropped. Only AI metrics with a minimum of 5 observations were included in our final dataset. For the datasets from Epoch AI, we analysed both the full datasets and a restricted dataset with data entries from January 1st, 2010, onwards.

We estimated the doubling time $T_d$ of each AI metric using the following method. For each AI metric of interest, we converted all dates to a number of days, where the time series start date was set to zero. We log-transformed all AI metric data (the dependent variables in our analysis) and fit a linear trendline using ordinary least squares (OLS) regression with the statsmodels package in Python. We extracted the regression coefficient $\hat{\beta}_1$, and estimated the doubling time as:

$$T_d = \frac{\ln 2}{\hat{\beta}_1}. \quad (1)$$

We estimated lower and upper confidence intervals (95%) of the doubling time by replacing $\hat{\beta}_1$ in Eq. 1 with the confidence interval bounds from the linear regression. When a categorical variable was available, we fit the data both with and without grouping the data by that categorical variable.

We filtered for fits fulfilling four conditions: $R^2 \geq 0.50$, $T_d > 0$, positive confidence intervals, and $CI_{upper} - CI_{lower} < 2\,T_d$, which returned 158 variables (out of our full database of 298). For the results from the Epoch AI dataset, we further filtered for estimates coming labelled as “Confident” under the categorical variable “Confidence” (i.e. discarding estimates from models with “Unknown”, “Speculative”, and “Likely” confidence). From the remaining fits, we selected a representative set of 12 variables to include in Fig. 1a of the main text (the exact values are presented in Supplementary Table 1).

**Supplementary Table 1: Doubling times of selected AI metrics.**

| AI metric | $T_d$ (months) | Lower CI (months) | Higher CI (months) | $R^2$ | N | Period | Source |
|---|---|---|---|---|---|---|---|
| Minimal model size (scoring >60% on MMLU) | 3.3 | 2.5 | 5.1 | 0.96 | 5 | 2022-04-04 to 2024-04-23 | AI Index Report (Fig. 2.1.38)[1,2] |
| Training compute of frontier AI models | 5.3 | 5.1 | 5.5 | 0.98 | 47 | 2010-09-26 to 2024-08-13 | Epoch AI (Data on AI Models)[3] |
| Task difficulty completed by AI (80% success) | 6.8 | 6.0 | 7.8 | 0.88 | 32 | 2019-02-14 to 2025-11-19 | METR[5,6] |
| Parameter size of frontier AI models | 8.9 | 7.20 | 11.6 | 0.66 | 40 | 2010-09-26 to 2024-07-23 | Epoch AI (Data on AI Models)[3] |
| Private investment in AI robotics globally | 9.5 | 6.1 | 21.9 | 0.80 | 7 | 2018-01-01 to 2024-01-01 | AI Index Report (Fig 4.3.16)[1,2] |
| NVIDIA compute installed globally | 9.8 | 9.5 | 10.1 | 1.00 | 24 | 2019-01-01 to 2024-10-01 | Epoch AI (NVIDIA chip production)[4] |
| Private investment in generative AI globally | 11.3 | 7.9 | 20.1 | 0.91 | 6 | 2019-01-01 to 2024-01-01 | AI Index Report (Fig. 4.3.3)[1,2] |
| New generative AI companies globally | 22.0 | 17.5 | 29.7 | 0.97 | 6 | 2019-01-01 to 2024-01-01 | AI Index Report (Fig. 4.3.5)[1,2] |
| Energy efficiency of leading AI hardware | 22.7 | 18.6 | 29.2 | 0.91 | 12 | 2016-04-05 to 2024-06-02 | AI Index Report (Fig. 1.4.5)[1,2] |
| Private investment in AI globally | 27.1 | 23.0 | 32.9 | 0.94 | 12 | 2013-01-01 to 2024-01-01 | AI Index Report (Fig. 4.3.2)[1,2] |
| AI patents granted globally | 30.4 | 26.7 | 35.3 | 0.95 | 14 | 2010-01-01 to 2023-01-01 | AI Index Report (Fig. 1.2.1)[1,2] |
| New industrial robots installed in China | 37.0 | 31.7 | 44.4 | 0.94 | 13 | 2011-01-01 to 2023-01-01 | AI Index Report (Fig. 4.5.5)[1,2] |

Note: CI = confidence interval. N = the number of points used in the regression. MMLU = Massive Multitask Language Understanding.

For Fig. 1a of the main text, doubling times were plotted in months, by dividing the estimated doubling time $T_d$ in days by 30.44 (the average number of days in a month within an average year of 365.25 days). AI metrics were split into three groups (Technical performance, Research and development, and Investment), following the approach used in the *AI Index Report*[2].

## Fig. 1b: Projection of AI capabilities

In Fig. 1b of the main text, we show the exponential trend of AI capabilities, and project this to the end of 2030. To construct the figure, we used data on the performance of GPT models over time, which we obtained from METR[5,6]. These data span the period from February 14th , 2019, until November 19th , 2025. We used the "p80_estimate" variable, which indicates "the length of tasks (measured by how long they take human professionals) that generalist autonomous frontier model agents can complete" with 80% probability[6]. We indexed values of the p80_estimate variable to the corresponding value for the "gpt_4" model.

For model fitting, we selected only the models from the METR dataset containing the string "gpt" (10 models out of 32), namely gpt-oss-120b, gpt2, gpt_3_5_turbo_instruct, gpt_4, gpt_4_0125, gpt_4_1106, gpt_4_turbo, gpt_4o, gpt_5, gpt_5_1_codex_max. We converted all dates to number of days, setting the start date to zero. We log-transformed the dependent variable (the indexed version of "p80_estimate") and then fit a linear trendline using OLS regression with the statsmodels package in Python.

For Fig. 1b of the main text, we have plotted the values for all 32 models contained in the original METR dataset, highlighting in red selected models.

## Fig. 2: AI futures and their implications for global GDP per capita

In Fig. 2 of the main text, we present a range of AI futures from the literature, and estimate global GDP per capita for each of them. Each AI scenario is calculated as a departure from the historical trend in global GDP per capita.

Global GDP data for the historical period from 1950 to 2022 were obtained by aggregating national data over this period. National GDP data were obtained from the 2023 version of the Maddison Project Database[7]. The variable that we used is "Real GDP per capita in 2011$". Since these data are in per capita terms, they were multiplied by population to give Real GDP in millions of 2011 US dollars.

For former Eastern Bloc countries within what used to be the Soviet Union, Yugoslavia, and Czechoslovakia, GDP data were not available for individual countries in the early years of the 1950–2022 period. However, the Maddison Project Database provides GDP estimates for the former Soviet Union, Yugoslavia, and Czechoslovakia back to 1950. National GDP estimates for the modern-day countries were therefore calculated by multiplying the regional totals for each year by the ratio of country-level GDP to regional GDP in the first year for which country-level data were available. Global annual GDP per capita data were then calculated by summing the GDP of all countries in each year, and dividing by global population. Population data were obtained from the United Nations World Population Prospects 2024 database[8], which provides annual population data for 1950–2024.

Given that the final year of data available from the Maddison Project Database dataset is 2022, global GDP values for 2023 and 2024 were obtained from Our World In Data (OWID)[9]. Since the OWID data are expressed in 2021 dollars, and the Maddison data in 2011 dollars, the OWID data were scaled to match the Maddison data in 2022 to ensure a consistent time series. As before, global GDP per capita values were calculated by dividing global GDP by global population.

Using the global GDP per capita dataset, the compound annual growth rate over the 1950–2024 period was calculated using log-linear regression, following the method suggested by Gujarati (pp. 169–171)[10]. Specifically, the growth rate of global GDP per capita was calculated as:

$$g = e^{\hat{\beta}_1} - 1 \tag{2}$$

where $\hat{\beta}_1$ is the slope of the best-fit line generated using OLS regression, after log-transforming the data. Using this method, we estimate an annual growth rate in global GDP per capita of $g$ = 2.07% for the 1950–2024 period.

The long-term trend in global GDP per capita over the full period from 1950 to 2100 period was then calculated as:

$$\hat{y}_t = e^{\hat{\beta}_0 + \hat{\beta}_1 t} \tag{3}$$

where $\hat{\beta}_0$ is the y-intercept of the OLS regression and $t$ is the year.

In addition to the long-term trend in global GDP per capita, Fig. 2 of the main text shows five scenarios of AI development: Intelligence explosion, Baseline AGI, Widespread adoption, Modest growth, and Human extinction. Each of these scenarios is based on a specific study published in the literature. Based on the description of the scenario in each study, we estimated a trajectory for global GDP per capita. Each scenario is briefly described below:

- **Intelligence explosion:** This scenario is based on Leopold Aschenbrenner's essay "Situational Awareness: The Decade Ahead"[11]. Aschenbrenner suggests that economic growth rates of 30% per year (or higher) could be reached by the end of the decade. For our version of this scenario, we assume that global GDP growth follows the historical trend ($g$ = 2.07% growth per year) until the end of 2026. In 2027 it increases to 15% per year, and from 2028 onwards it skyrockets to 30% per year.

- **Baseline AGI:** This scenario is based on the baseline AGI scenario from Korinek and Suh's analysis[12], in which full automation occurs over a 20-year period. In this scenario, output growth increases from approximately 2% per year to 18% per year. In our version of this scenario, global GDP growth follows the historical trend until the end of 2025. Over the 20-year period from 2026 until 2045, the growth rate increases linearly from 2.07% per year until it reaches 18% per year. Growth then continues at this rate from 2045 onwards.

- **Widespread adoption:** This scenario is based on a study by Goldman Sachs[13], which estimates that widespread AI adoption could increase global GDP by 7% over a 10-year period, compared to what it would otherwise be. A 7% increase over 10 years corresponds to a growth rate of 0.68% per year. In our version of this scenario, global GDP growth follows the historical trend until the end of 2025. Over the 10-year period from 2026 to 2035, it grows by $g$ + 0.68% = 2.74% per year. From 2036 onwards, the growth rate reverts to the historical trend of 2.07% per year.

- **Modest growth:** This scenario is based on a study by Acemoglu[14], which estimates that the boost to GDP from AI will be modest. He estimates that US GDP could increase by 0.93–1.16% over 10 years in total (with modest investment), or by 1.40–1.56% in total (with a large investment boom). For our scenario, we take the midpoint of these estimates (1.25% growth in total over 10 years), which corresponds to a growth rate of 0.12% per year. In our version of this scenario, global GDP growth follows the historical trend until the end of 2025. From 2026 onwards it grows by $g$ + 0.12% = 2.19% per year.

- **Human extinction:** In their book, *If Anyone Builds It, Everyone Dies,* Yudkowsky and Soares argue that the creation of superintelligent AI would likely to lead to human extinction[15]. With respect to economic growth, Yudkowsky has stated, "My mainline view is that growth stays at 5%/year and then everybody falls over dead in 3 seconds and the world gets transformed into paperclips"[16]. For our version of this scenario, we assume that global GDP growth follows the same trajectory as the baseline AGI scenario from 2025 until 2030. In 2031, in plummets abruptly to zero.

The five scenarios are summarised in Supplementary Table 2.

**Supplementary Table 2: Five AI development scenarios.**

| Scenario | Study | Description from study of what happen to GDP | Timeframe | Annualised growth rate |
|---|---|---|---|---|
| Intelligence explosion | Aschenbrenner[11] | "We could see economic growth rates of 30%/year and beyond, quite possibly multiple doublings a year" (p. 69); "It is plausible growth rates could go into the 10s of percent a year" (p. 129). | By the end of the decade | 30% |
| Baseline AGI | Korinek and Suh[12] | "In the baseline AGI scenario, wages grow slightly during the initial periods but then collapse before full automation is reached. After the collapse, wages are equal to the returns to capital, ... labor and capital are perfectly substitutable, with steady-state growth of 18% per year" (p. 25). | 20-year period | 18% |
| Widespread adoption | Goldman Sachs[13] | "Applying our estimated global labor productivity boost to countries in our coverage implies that widespread AI adoption could eventually drive a 7% or almost $7tn increase in annual global GDP over a 10-year period" (pp. 15-16). | 10-year period | $g$ + 0.68% = 2.74% |
| Modest growth | Acemoglu[14] | "[M]y calculations suggest that the GDP boost within the next 10 years should also be modest, in the range of 0.93%–1.16% over 10 years in total, provided that the investment increase resulting from AI is modest, and in the range of 1.4%–1.56% in total, if there is a large investment boom" (p. 54). | 10-year period | $g$ + 0.12% = 2.19% |
| Human extinction | Yudkowsky and Soares[15,16] | "My mainline view is that growth stays at 5%/year and then everybody falls over dead in 3 seconds and the world gets transformed into paperclips". | Very fast once it starts | -99.9% |

## Fig. 3: Job satisfaction and meaning in relation to occupational exposure to automation

In Fig. 3 of the main text, we compare the job satisfaction and meaning that people experience from different occupations to how exposed these occupations are to computer automation. Using estimates of exposure to automation from two different studies spaced ten years apart, we show that while jobs that were estimated to be more automatable in the past were also those with lower job satisfaction and meaning, this relationship does not hold in the most-recent data. Large language models (LLMs) are now just as able to automate jobs with high satisfaction and meaning as jobs with low satisfaction and meaning.

We obtained job satisfaction and meaning data from a survey conducted by Payscale[17]. The survey was completed by 2.7 million people over a two-year period between 11 June 2013 and 11 June 2015. The survey provides results for 505 occupations, following standard categories published by the Occupational Information Network (O*NET).

We used two questions from the survey: (1) percentage of workers reporting high satisfaction with their job (specifically the percentage of respondents with the given job who answered "Extremely satisfied" or "Fairly satisfied" to the question, "How satisfied are you in your job?"), and (2) percentage of workers reporting that their job was meaningful (specifically, the percentage of respondents with the given job who answered "Very much so" or "Yes" to the question, "Does your work make the world a better place?"). The results for these two questions were averaged to create a single index of job satisfaction and meaning for each occupation.

Exposure to automation data were obtained from two different sources. The first source is a well-known study by Frey and Osborne, which was first published in 2013 as a working paper[18], and then re-published in 2017 as a journal article[19]. The results presented in the two versions are the same. They estimate the probability of computerisation for 702 occupations, classified according to six-digit Standard Occupational Classification (SOC) codes following the O*NET SOC 2010 system.

The second source of automation data that we used is a more recent study by Eloundou et al., which was first published in 2023 as an OpenAI report[20], and then in 2024 as a journal article[21]. We obtained the full dataset from the GitHub repository that accompanies the article[22]. The study estimates the automation exposure of 923 occupations, with occupations classified according to eight-digit SOC codes using the O*NET SOC 2019 system. The authors define exposure as "the capacity of an LLM or LLM-powered system to reduce the time required for a human to complete a task by at least 50% while preserving or improving quality"[21]. The authors present both human ratings and GPT-4 ratings for the occupations. For our analysis we use the human ratings, specifically the indicator $\zeta = E1 + E2$, which represents the share of tasks exposed to automation assuming full LLM and software integration.

To combine the three datasets, we matched all data to the six-digit O*NET SOC 2010 codes published by the US Bureau of Labor Statistics[23]. These are the codes used by Frey and Osborne, so no additional processing was needed for these data. The Payscale data provide occupation descriptions based on the same O*NET SOC 2010 system, which were matched to the corresponding six-digit codes. The data provided by Eloundou et al. use the newer O*NET SOC 2019 system, which consists of eight-digit codes and a larger number of occupations. The Eloundou et al. data were mapped to the level of six-digit codes by taking the average of the automation exposure values for all eight-digit occupations falling under each six-digit code. After combining the three datasets, we were left with 417 occupations for which all variables were available.

We used OLS regression to test the dependence of our index of job satisfaction and meaning on the two estimates of exposure to automation. The regression results are shown in Supplementary Table 3, and plotted in Fig. 3 of the main text. The results show that occupations with higher exposure to automation, as estimated by Frey and Osborne in 2013, have lower job satisfaction and meaning, and this result is highly statistically significant ($p < 0.001$). These results are consistent with previous findings in the literature[24,25]. However, with the more recent estimates of exposure to automation by Eloundou et al. from 2023, the relationship largely disappears. The newer data suggest that occupations with higher exposure to automation actually have higher job satisfaction and meaning, although this relationship is weak, and only marginally statistically significant ($p < 0.05$).

**Supplementary Table 3: OLS regression results for job satisfaction and meaning.**

| Variable | Model 1 | | Model 2 | |
|---|---|---|---|---|
| Frey and Osborne's estimate of exposure to automation (2013) | -0.166 | *** | | |
| | (0.013) | | | |
| Eloundou et al.'s estimate of exposure to automation (2023) | | | 0.046 | * |
| | | | (0.020) | |
| Constant (Intercept) | 0.699 | *** | 0.597 | *** |
| | (0.008) | | (0.011) | |
| Observations (N) | 417 | | 417 | |
| Adjusted R-Squared | 0.269 | | 0.010 | |

Note: Standard errors are reported in parentheses. * $p < 0.05$, ** $p < 0.01$, *** $p < 0.001$

## Fig. 4: Applying the Doughnut of social and planetary boundaries to assess and guide AI development

In Fig. 4 of the main text, we show how the Doughnut of social and planetary boundaries could be applied to assess and guide AI development. The figure overlays two published scenarios on top of each other, to show how they differ. The first scenario shows the actual global data for the year 2022, as published by Fanning and Raworth[26]. In cases where the original study by Fanning and Raworth presented more than one indicator for a given social or environmental dimension, the indicators have been averaged to show only one value for each dimension.

The second scenario is a hypothetical AI development scenario from O'Neill and Creutzig[27]. In this scenario, new technologies improve four of the social dimensions of the Doughnut (health, education, energy access, and connectivity), but the unequal distribution of the benefits from AI worsens three of the other social dimensions (employment, equality, and social cohesion). On the environmental side, precision agriculture reduces land conversion and nutrient pollution, improving these indicators, but the expansion of large data centres increases water use and $CO_2$ emissions, worsening climate change.

## Supplementary References